\documentclass[aps,article,prc,nofootinbib,preprintnumbers,amsmath,amssymb,twocolumn,superscriptaddress,showpacs,floatfix]{revtex4}

\usepackage{epsfig}
\usepackage{dcolumn}
\usepackage{bm}
\newcommand{\order}[1]{ \mathcal{O} \left( #1 \right) }

\newcommand{\ave}[1]{\left\langle #1 \right\rangle}

\begin{document}
\title{Fourier Harmonics of High-$p_T$ Particles Probing the Fluctuating Inititial Condition Geometries in
Heavy-Ion Collisions}

\author{Barbara Betz}
\affiliation{Department of Physics, Columbia University, 
New York, 10027, USA}
\author{Miklos Gyulassy}
\affiliation{Department of Physics, Columbia University, 
New York, 10027, USA}
\author{Giorgio Torrieri}
\affiliation{Frankfurt Institute for Advanced Studies (FIAS), 
Frankfurt am Main, Germany}

\begin{abstract}

Second Fourier harmonics of jet quenching have been thoroughly explored 
in the literature and shown to be sensitive to the underlying jet 
path-length dependence of energy loss and the differences between the 
mean eccentricity predicted by Glauber and CGC/KLN models of initial 
conditions. We compute the jet path-length dependence of energy-loss
for higher azimuthal harmonics of jet-fragments in a generalized
model of energy-loss for RHIC energies and find, however, that even the 
high-$p_T$ second moment is most sensitive to the poorly known early-time 
evolution during the first fm/c. Moreover, we demonstrate that higher-jet 
harmonics are remarkably insensitive to the initial conditions, while the 
different $v_n(N_{part})$ vs.\ $v_n^{I_{AA}}(N_{part})$ correlations 
between the moments of monojet and dijet nuclear modifications factors 
remain a most sensitive probe to differentiate between Glauber and 
CGC/KLN initial state sQGP geometries. 

\end{abstract}

\maketitle

\section{Introduction}

Heavy-ion collisions at the Relativistc Heavy Ion Collider (RHIC)
indicate the production of an opaque (i.e.\ strongly jet-suppressing)
\cite{sqgp,whitebrahms,whitephobos,whitestar,whitephenix},
strongly-coupled, fast-thermalizing medium that possibly needs
to be described using methods derived from AdS/CFT string theory
\cite{coester}. However, so far neither the initial conditions of the 
collisions nor the microscopic dynamics of the jet-energy loss are 
conclusively understood. 

Two models are commonly used to characterize the initial conditions. 
The Glauber model \cite{glauber}, describing incoherent superpositions 
of proton-proton collisions, and the ``Color Glass Condensate'' (CGC) 
\cite{cgc}, given e.g.\ by the KLN model \cite{kln,adrian,dumitru}, 
where saturation effects are taken into account. They differ by their
initial temperature gradients, their initial high-$p_T$ parton distribution, 
and the distance travelled by each parton, leading to a different opacity 
estimate.   
 In addition, both models exhibit large event-by-event fluctuations 
\cite{sorfluct,phogla,grassi,alver}.

The jet-energy loss can either be described as multiple scatterings of
the parton \cite{bh,lpm,Baier:1996sk,Wiedemann:2000za,Qiu:1990xa,Arnold:2001ba}, 
specific for a weakly-coupled pQCD medium, or using the AdS/CFT correspondence 
where the problem of a parton stopped in a thermal medium is related to the 
problem of a string falling into a $5$-dimensional black hole 
\cite{ches1,ches2,ches3,Arnold:2011qi}.

Experimentally, jet-energy loss is parametrized by the suppression factor $R_{AA}$,
defined as the ratio of jets produced in $A+A$ collisions to the expectation 
for jets being produced in $p+p$ collisions
\begin{equation}
 R_{AA}(p_T) = \frac{dN^{A+A}/dp_T}{N_{coll}dN^{p+p}/dp_T}\;,
\end{equation}
where $N_{coll}$ is the number of collisions, a theoretical parameter 
(calculated within the Glauber model \cite{phogla}) depending on centrality 
(i.e., on the number of participants $N_{part}$).

The first attempt to disentagle the initial conditions (Glauber vs.\ CGC) 
and the energy-loss mechanism (pQCD vs.\ AdS/CFT) while simultaneously 
describing the nuclear modification factor $R_{AA}(N_{part})$
and the elliptic flow $v_2(N_{part})$ for high-$p_T$ particles 
was given in Refs.\ \cite{Jia,Adare:2010sp,horojia}, favoring CGC initial conditions 
and a strongly-coupled energy loss at RHIC. A similar ansatz was used in 
Ref.\ \cite{Fries:2010jd}, although the Fourier components were not
shown explicitly. 

To further study the differences of a pQCD and an AdS/CFT-like energy loss,
we investigate the role of the energy dependence in the energy-loss 
prescription, examine the power of the path-length dependence, and calculate
higher-jet harmonics. We want to examine if a generic energy-loss ansatz
that includes both a path-length and an energy dependence confirms the 
above conclusion that only CGC initial conditions and an AdS/CFT 
energy-loss can describe both the $R_{AA} (N_{part})$ and
the $v_2 (N_{part})$ appropriately.

In the high-temperature limit, all dependences on the intrinsic scales 
of the system ($T_c, \Lambda_{QCD}$, etc.) disappear. Because of this, a 
generic energy-loss rate $dE/dx$ is given by an arbitrary combination 
of dimensionful parameters constrained by the total dimension of the 
observable and the requirement that faster particles and hotter media 
result in a bigger suppression. We choose
\begin{eqnarray}
\hspace*{-0.4cm}
\frac{dE}{dx}(\vec{x}_0,\phi,\tau)&=&
-\kappa P^a\tau^{z} T^{z-a+2}[\vec{x}_0+\hat{n}(\phi)\tau,\tau],
\label{GenericEloss}
\end{eqnarray}
for an energy loss as a function of time $\tau$ considering a jet 
starting at $\vec{x}_0$ and propagating in direction $\phi$ with the 
coupling $\kappa$. $P$ is the momentum of the jet(s) considered, $T$ is
the temperature, and $a,z$ are parameters 
controlling the jet energy (momentum) and path-length dependence, respectively. 

In the Bethe-Heitler limit $a=1$ and $z=0$, while in the deep LPM pQCD 
limit $a\sim0$ and $z\sim 1$. 
If $a=0$ and $z=2$, our model coincides with the model referred to as "AdS/CFT" in 
Refs.\ \cite{Jia,horojia}. However, on-shell AdS/CFT calculations 
\cite{ches1,ches2,ches3,Arnold:2011qi} show that $a=1/3$ and $z=2$ , 
thus we are going to consider $a=1/3$ throughout the whole paper. 
However, one should keep in mind that $a=1/3$ is a weak lower bound
for falling strings used for illustration. 
We note that, as it is clearly shown in Refs.\
\cite{ches1,ches2,ches3,Arnold:2011qi}, $z=2$ is only a {\it lower} 
limit corresponding to an ``on-shell'' quark whose stopping 
distance is $l_f \gg 1/T$ and whose initial energy is
$E_0 \gg T$. In a realistic medium, the first assumption is likely to be 
violated, resulting in values possibly of $z>2$.

Please note that in contrast to Refs.\ \cite{Jia,Adare:2010sp,horojia}, 
$\kappa$ is a dimensionless parameter.
In a particular case of radiative-dominated scattering in the LPM regime, 
$\kappa T^3 \sim \hat{q}$ \cite{qhat}.  In general, $\kappa$ measures 
the interaction cross-section, while $T^3$ is related to the density of 
scattering centers. This (like soft observables in general) can be used 
as a constraint for $\kappa$, but additionally one has to assume nearly 
complete thermalization on the timescale of jet propagation throughout 
the system as well as a straight-forward relation between entropy density 
and multiplicity. While these assumptions are reasonable, they are not 
easily falsifiable, and, in particular, in the $\hat{q}$-limit, one finds 
that $\kappa$ can {\em not} describe jet quenching and a realistic 
gluon density at the same time \cite{qhat}. Therefore, we limit ourselves 
to fitting $\kappa$ to the most central data point of $R_{AA}$ measured 
at RHIC \cite{Adare:2010sp}, disregarding any interpretation in terms 
of the density of soft degrees of freedom, comparing to the RHIC data on
$(\vec{x}_0,\phi)$-averaged $R_{AA}$ versus centrality for a range of 
$E_f\sim 6-9$ GeV \cite{Adare:2010sp}.

In a static medium, $dE/dx \sim \tau^z$, while in a dynamic medium, $dE/dx$ will 
aquire additional powers of $\tau$ due to the dependence of temperature
on $\tau$, implicitly included in Eq.\ (\ref{GenericEloss}).  
Here, we assume a $1$D Bjorken expansion
\cite{bjorken} 
\begin{equation}
T(\vec{x},\tau) = T_0(\vec{x}) \left( \frac{\tau_0}{\tau} \right)^{1/3}
\end{equation}
with different values of $\tau_0$. Surprisingly enough, we find that
even the high-$p_T$ second Fourier moment ($v_2$) is most 
sensitive to the poorly known early-time evolution during the first fm/c.
 
Considering that $R_{AA}$ is given by the ratio of jets in the QGP
to jets in vacuum, one obtains the following formula for the nuclear
modification factor from Eq.\ (\ref{GenericEloss}) for a jet starting at
$\vec{x}_0$ and propagating in direction $\phi$
\begin{equation}
R_{AA}(N_{part},\vec{x}_0,\phi)= \exp[-\chi(\vec{x}_0,\phi)]\,,
\end{equation}
with
\begin{equation}\hspace*{-1ex}
\chi(\vec{x}_0,\phi) = 
\left(\frac{1+a-n}{1-a}\right)\ln\left[ 1- 
\frac{K}{P_0^{1-a}}I(\vec{x}_0,\phi,a,z) \right],
\end{equation}
where $n$ is the spectral index (taken to be $n\sim 6$), $K = \kappa (1-a)$, 
$P_0$ is the jet's initial momentum and the line integral is
\begin{equation}
I(\vec{x}_0,\phi,a,z)=\int\limits_{\tau_{0}}^\infty 
\tau^{z} T^{z-a+2}[\vec{x}_0+\hat{n}(\phi)\tau,\tau]d\tau\,.
\label{OurLineint}
\end{equation}

In case of $a=1$ (the Bethe-Heitler limit), $1/(1-a)$ diverges, but since
$K=\kappa(1-a) \rightarrow 0$, 
\begin{eqnarray}
\hspace*{-2ex}
\chi(\vec{x}_0,\phi) &=& \kappa (n-2) I(\vec{x}_0,\phi,a=1,z).
\end{eqnarray}

From the above calculated $R_{AA}(N_{part},\vec{x}_0,\phi)$, the $R_{AA}(N_{part})$ 
can be obtained by averaging over all possible $\vec{x}_0$ and $\phi$
\begin{eqnarray}
R_{AA}(N_{part})&=&\int\, \frac{d\phi}{2\pi} \frac{\int R_{AA}(N_{part},\vec{x}_0,\phi)\, T_{AA}(\vec{x}_0) d\vec{x}_0}
{\int T_{AA}(\vec{x}_0) d\vec{x}_0}\nonumber\\
&=&\int  \frac{d\phi}{2\pi} R_{AA}(N_{part},\phi)\,,
\end{eqnarray}
where the nuclear overlap function is used in case of the Glauber model and
$T_{AA}=\rho^2/P_0^2$ for the CGC prescription.

After having fixed $\kappa$, the $v_n(N_{part})$ can be computed via
\begin{eqnarray}
\hspace*{-2ex}
v_n(N_{part})
&=&\frac{\int d\phi \cos\left\{n\left[ \phi-\psi_n \right]\right\} \,R_{AA}(N_{part},\phi)}
{\int d\phi \,R_{AA}(N_{part},\phi)} \,.
\end{eqnarray}
The Fourier density components $e_n$ and the reaction-plane axis $\psi_n$ are 
determined according to the initial density distribution used in Ref.\
 \cite{alver}
\begin{eqnarray}
e_n (t) &=& \frac{\sqrt{\ave{ r^2 \cos(n\phi)}^2 + \ave{ r^2 \sin(n\phi) }^2}}{\ave{ r^2 }}
\end{eqnarray}
and
\begin{equation}
\psi_n (t) = \frac{1}{n}\tan^{-1} \frac{\ave{ r^2 \sin(n\phi)}}{\ave{ r^2 \cos(n\phi)}}\,.
\end{equation}
By definition, the impact parameter points 
into the $x$-direction, but event-by-event fluctuations introduce 
non-trivial and harmonic dependent $\psi_n$'s.

We checked that we reproduce the results of Ref.\ \cite{Jia} when our 
approach is simplified to the one used in this reference, where
\begin{eqnarray}
R_{AA}(N_{part})&=&\langle e^{-\kappa I_m}\rangle\,,
\end{eqnarray}
and the line integral is
\begin{equation}
I_m = \int_0^\infty dl\, l^{m-1} \, \rho(\vec{r}+l\hat{v}), m=1,2,...
\label{lineintJia}
\end{equation}
Here, $m=1$ and $m=2$ describe the path-length ($l$) dependence for 
pQCD and AdS/CFT-like energy-loss, repsectively.
However, one of the main differences to our approach is that we also 
consider the energy dependence in the jet-energy loss. In other words, 
$a=0$ in Refs.\ \cite{Jia,Adare:2010sp,horojia} while we assume $a=1/3$ [see Eq.\ 
(\ref{GenericEloss})] throughout the whole study.

The above analysis can be extended to dijets, analogously to Ref.\ \cite{horojia}. 
Dijet suppression is parametrized by the factor $I_{AA}$, the ratio of the 
dijet suppression to the suppression of jets,
 defined experimentally as in \cite{iaaref} and related to the parameters we defined earlier as in \cite{horojia}
\begin{eqnarray}
\hspace*{-2ex}
I_{AA} &=&  \frac{dN_{dijet}^{A+A}/dp_T}{R_{AA} dN_{dijet}^{p+p}/dp_T}
\backsimeq \frac{\langle e^{-\kappa(I^t + I^a)}\rangle}{R_{AA}} \,,
\end{eqnarray}
with the line integrals for the trigger (t) and away-side (a) jet. 
Please note that in contrast to Ref.\ \cite{horojia},
our coupling $\kappa$ is equal for the trigger and the away-side jet.
Clearly, a higher $I_{AA}$ then $R_{AA}$, as observed in Ref.\ \cite{iaaref}, 
implies that if one jet survives, the other one has a higher probability 
to survive as well. In turn, this indicates that jets emitted from 
less dense periphery regions have a larger impact. Thus, as remarked in 
Ref.\ \cite{iaaref}, comparing jet abundance to dijet abundance can be a 
sensitive medium probe.

\section{Results and discussion}
\begin{figure}
\hspace*{-0.6cm}
\includegraphics[scale=0.5]{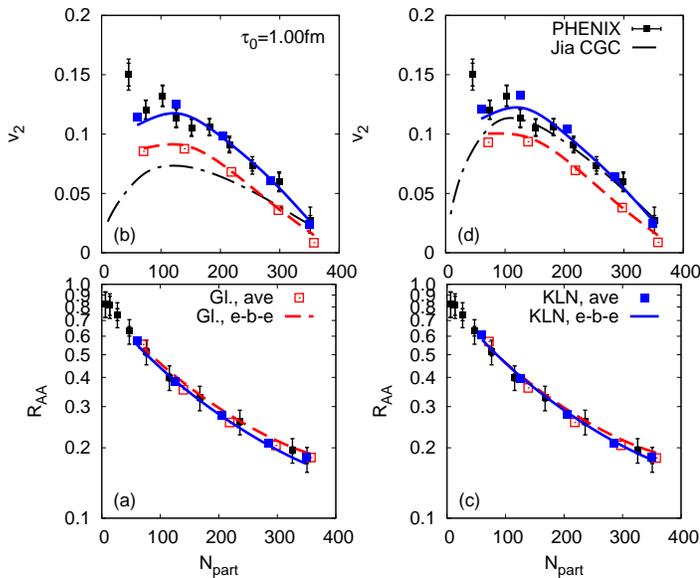}
\caption{(Color online) $v_2$ (top panels) and $R_{AA}$ (bottom panels) of high-momentum 
particles as a function of the number of participants, $N_{part}$, for 
$z=1$ (left panel) and $z=2$ [right panel, 
see Eq.\ (\ref{GenericEloss}) for definition] and $\tau_0=1$~fm. 
The RHIC data are taken from Ref.\ \cite{Adare:2010sp}.
\label{v2Raa}}
\end{figure}
\begin{figure}
\hspace*{-0.6cm}
\includegraphics[scale=0.5]{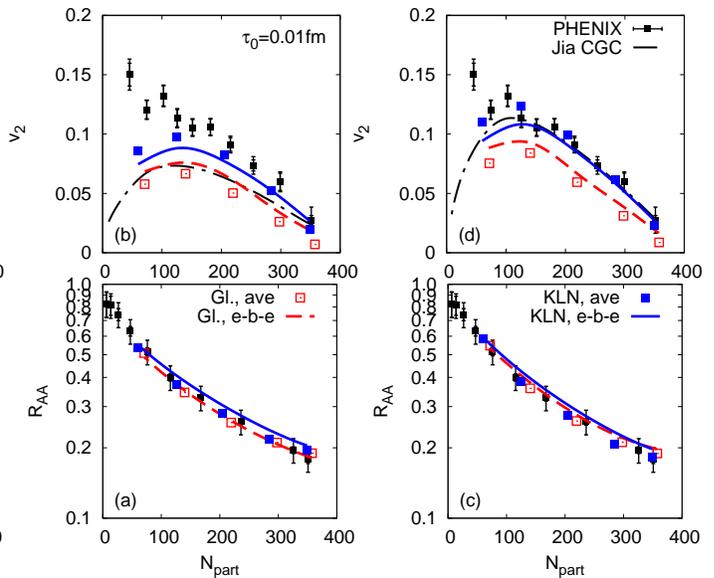}
\caption{(Color online)  $v_2$ (top panels) and $R_{AA}$ (bottom panels) of 
high-momentum particles as in Fig.\ \ref{v2Raa} for $\tau_0=0.01$~fm.}\label{v2t0}
\end{figure}
\begin{figure}
\hspace*{-0.6cm}
\includegraphics[scale=0.5]{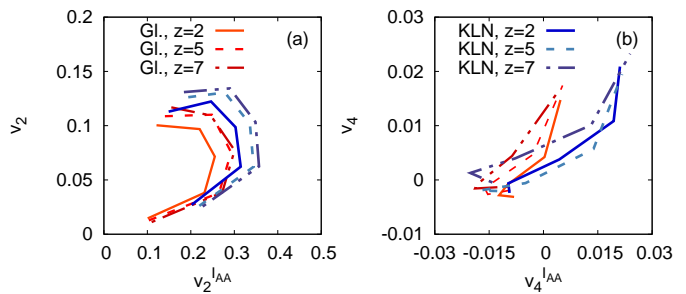}
\caption{(Color online) Higher Fourier coefficients of dijet observables, 
$v_2^{I_{AA}}$ vs.\ $v_2$ (a) and $v_4^{I_{AA}}$ vs.\ $v_4$
(b) for the Glauber model (reddish lines) and the KLN
model (bluish lines) for $z=2,5,7$ and $\tau_0=1$~fm. A clear shift between
the Glauber and the KLN model as well as a saturation effect for larger 
$z$ can be seen.
\label{vnvnIAA}}
\end{figure}

\subsection{Mean Correlations}
We have carried out the procedure described in the previous section for a variety 
of impact parameters and initial conditions at RHIC energies. LHC energies
have been mentioned in Ref.\ \cite{QMProceedings,Horowitz:2011gd} and will 
be explored in detail in Ref.\ \cite{UsFuture}. 
In this subsection, we focus on the mean correlations. The width of 
these correlations will be discussed in the following subsection. Please 
note that this width is the actual physical geometry fluctuation.

Throughout the paper, we distinguish four different cases. The Glauber model
\cite{glauber} and the CGC/KLN prescription \cite{cgc,kln,adrian}, both
event-by-event and averaged over many events, i.e.\ in the latter case 
the initial conditions are smoothed over many events and the energy loss 
is calculated subsequently.

In Fig.\ \ref{v2Raa} we choose $\tau_0=1$~fm, in line with recent hydrodynamic 
calculations \cite{heinzrecent}.
The figure shows that for both Glauber (red) and KLN (blue) initial conditions
as well as for both pQCD-like [Fig.\ \ref{v2Raa} (a)] and 
AdS/CFT-like [Fig.\ \ref{v2Raa}(c)] energy loss the $R_{AA}(N_{part})$ 
can be described choosing an appropriate value for $\kappa$.  
However, surprisingly enough, the results for $v_2 (N_{part})$ get close to
the RHIC data when using KLN initial conditions 
for both pQCD-like [Fig.\ \ref{v2Raa}(b)] and AdS/CFT-like 
[Fig.\ \ref{v2Raa}(d)] energy loss, while Glauber 
initial conditions underpredict the data. Moreover, the difference between
pQCD-like and AdS/CFT-like energy loss is rather weak. 

This result is a clear contradiction to the one shown in Ref.\ \cite{Adare:2010sp}
and questions the conclusion that only CGC/KLN initial conditions and an AdS/CFT 
energy-loss can describe both the $R_{AA} (N_{part})$ and
the $v_2 (N_{part})$ appropriately.
However, choosing a much smaller $\tau_0$, as done in Fig.\ \ref{v2t0},
reduces the absolute value of $v_2 (N_{part})$ for both pQCD and 
AdS/CFT-like energy loss scenarios, while it increases the difference between the
pQCD and AdS/CFT results as seen in Refs.\ \cite{Jia,Adare:2010sp}. 
Here, the difference to the fit by Jia et al.\ \cite{Jia}
(black long dashed-dotted line), mainly seen for the pQCD-like energy-loss, 
is due to the additional energy loss dependence, parametrized by $a=1/3$.
Additionally, both plots show that there is a small discrepancy between the
event-by-event and the averaged scenario \cite{Renk:2011qi}.

We would like to mention here that in Refs.\ \cite{Jia} different values 
of $\tau_0$ were analyzed, nevertheless only the $\tau_0=0$~fm case
was shown in Ref.\ \cite{Adare:2010sp}, leaving out a discussion about
the physical meaning of $\tau_0$.

Setting $\tau_0=1$~fm means to assume that there is no energy loss 
within the first fm. PQCD does not give any excuse for this assumption and
thus $\tau_0=0$~fm would be a natural assumption. However, $\tau_0$
also describes the formation time of hydrodynamics which seems to be
$\tau_0\sim 1$~fm \cite{heinzrecent}.
On the other hand, setting $\tau_0=1$~fm is also equivalent to the 
AdS/CFT result that the energy loss is suppressed at early times (due
to the $dE/dx\sim l^2$ dependence). Thus, it is important to note that
the $v_2$ of high-$p_T$ particles is sensitive to short-distance properties,
suggesting that there is either weak coupling with a $\tau_0\sim 1$~fm or
strong coupling which in itself features the suppression of energy loss
at early times.

Extending the analysis to dijets and calculating the $I_{AA}$, our results meet
the datapoint of Ref.\ \cite{horojiadata} for both Glauber and KLN initial 
conditions when assuming that $\kappa$ is the same for both parts of the dijet, 
in contrast to Ref.\ \cite{horojia}. 

Since the energy loss calculated by AdS/CFT \cite{ches1,ches2,ches3,Arnold:2011qi}
clearly states that $z=2$ is only a lower limit, this raises the question 
which power in the jet path-length dependece has to be included when
an AdS/CFT-like energy loss is considered. A detailed discussion of this issue 
will follow in the next subsection (see the Fig.\ \ref{expfig}).

Before that, we will focus on correlations of higher Fourier coefficients
of dijet observables, $v_2^{I_{AA}}$ vs.\ $v_2$ and $v_4^{I_{AA}}$ vs.\ $v_4$.
Fig.\ \ref{vnvnIAA} displays those correlations for the Glauber (reddish
lines) and the KLN model (bluish lines) for $z=2,5,7$.

A clear saturation effect occurs for larger $z$ as well as a shift 
between the Glauber and the KLN model. Please note that while for the
$v_2$ correlation larger values of $z$ of the Glauber model coincide with 
lower values of the KLN model, this is no longer true for the $v_4$ correlations.
Therefore, the different means of the correlations between $v_n^{I_{AA}}(p_T)$ vs.\ $v_n(p_T)$ 
remain a most sensitive probe to differentiate between CGC/KLN and Glauber 
initial state geometries.

\subsection{The Shallowness of the Correlations}

After having focussed on the mean correlations in the last subsection, 
we now want to investigate the width of these correlations which is 
extremely important to conclude about the significance of the previously 
shown differences in the path-length dependence of the energy loss 
and the initial states considered.

\begin{figure}
\hspace*{-0.6cm}
\includegraphics[scale=0.5]{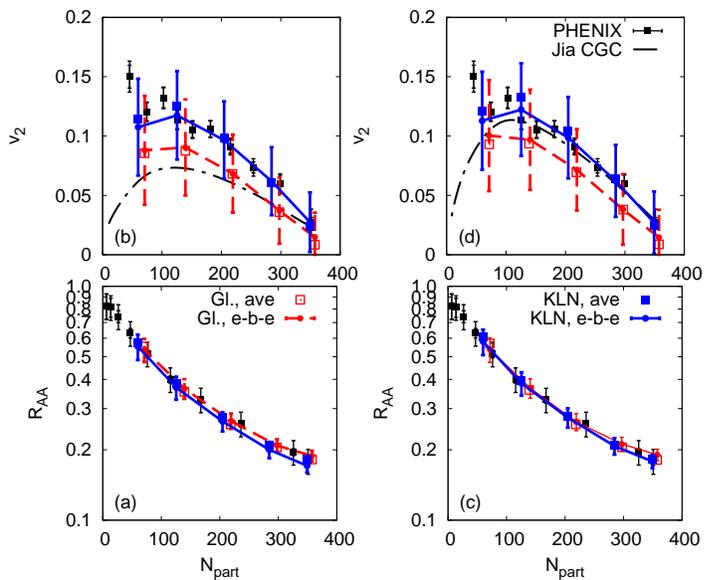}
\caption{(Color online) The fluctuation of $v_2$ (top panels) and $R_{AA}$ (bottom panels) of high-momentum 
particles as a function of the number of participants, $N_{part}$, 
for $z=1$ (left panel) and $z=2$ (right panel) and $\tau_0=1$~fm. The RHIC 
data are taken from Ref.\ \cite{Adare:2010sp}.
\label{v2fig}}
\end{figure}
\begin{figure}
\hspace*{-0.6cm}
\includegraphics[scale=0.5]{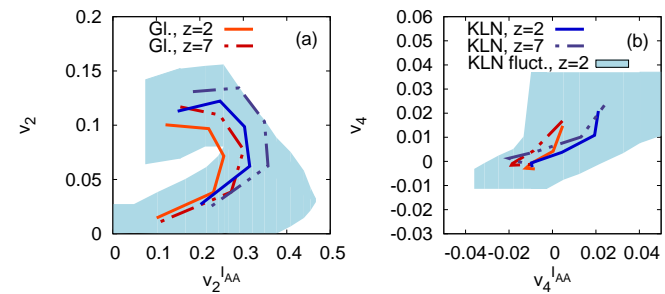}
\caption{(Color online) The fluctuation of the higher Fourier coefficients of dijet observables, 
$v_2^{I_{AA}}$ vs.\ $v_2$ (a) and $v_4^{I_{AA}}$ vs.\ $v_4$
(b) for the Glauber model (reddish lines) and the KLN
model (bluish lines) for $z=2,7$ and $\tau_0=1$~fm. The blue area displays the region
covered by the error bars of the KLN scenario for $z=2$. The other
cases show similar widths.
\label{iaafig}}
\end{figure}

Figs.\ \ref{v2fig} and \ref{iaafig} are a repetition of Figs.\ \ref{v2Raa} 
and \ref{vnvnIAA}, including the fluctuations for the event-by-event
analysis. As can be seen, those fluctuations are small for $R_{AA}(N_{part})$,
but the width for the $v_2(N_{part})$ is rather large. Thus, in an 
experimental analysis it is less straighforward to distinguish
between the different initial conditions than originally hoped for.
Nevertheless, the CGC/KLN initial conditions seem to be favorable.

In Fig.\ \ref{iaafig} it becomes obvious that the while larger 
Fourier coefficients become smaller, their width becomes larger, 
making it experimentally more difficult to disentangle the different
scenarios. Please note that the scale of Fig.\ \ref{iaafig}(b) 
is larger than the one of Fig.\ \ref{vnvnIAA}(b),
emphasizing the very broad event-by-event fluctuations of the 
higher-moment correlations. Therefore, in order to draw an experimentally testable
conclusion, it is necessary to always determine the mean and the width 
of the correlations considered.

While the 2nd Fourier harmonics of jet quenching have been thoroughly investigated
\cite{Jia,horojia,renk}, the sensitivity of higher harmonics has 
remained relatively unexplored. In Ref.\ \cite{Fries:2010jd}
higher harmonics were mentioned but not shown explicitly.

\begin{figure}
\hspace*{-0.6cm}
\includegraphics[scale=0.5]{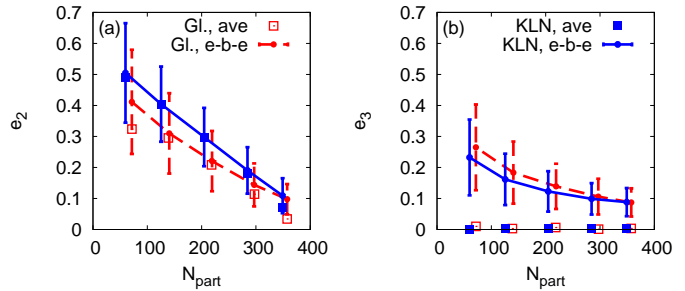}
\caption{(Color online) The eccentricity of the second (a) and third (b) Fourier components
of the Glauber and KLN model, both event-by-event (red and blue lines)
and averaged (red and blue dots). Since $e_3$ is only present when
fluctuations are taken into account, it has to vanish for the averaged
analysis.
\label{efig}}
\end{figure}

In a first step, we examine the differences in the eccentricities
between the Glauber and the KLN model, see Figure \ref{efig}. Here it
becomes clear again that while on average differences certainly exist
(that are bigger for $e_2$ than for $e_3$), the width of the distributions 
is again rather large. This is true both for harmonics present 
on average as well as event-by-event, such as $e_2$, and 
harmonics only present once fluctuations are taken into account, such as 
$e_3$, suggesting that higher harmonics of jet observables are rather
insensitive to initial conditions.

\begin{figure}[b]
\hspace*{-0.6cm}
\includegraphics[scale=0.5]{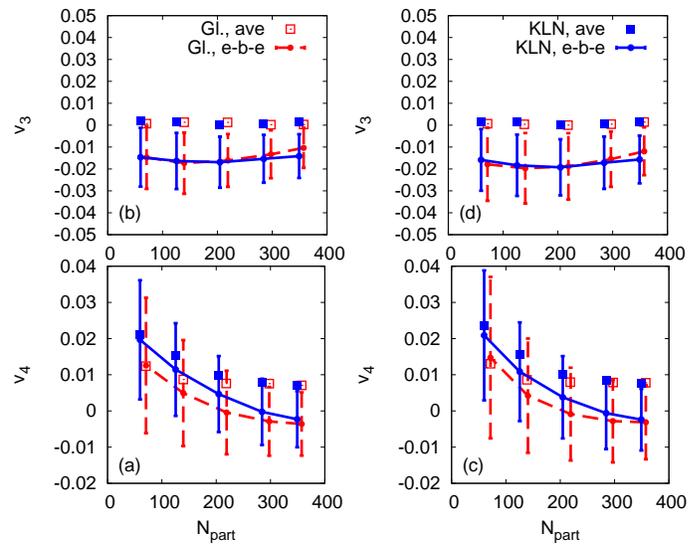}
\caption{(Color online) $v_3$ (top panels) and $v_4$ (bottom panels) of high-momentum 
particles as a function of the number of participants, $N_{part}$, for 
$z=1$ (left panel) and $z=2$ (right panel) and $\tau_0=1$~fm.
\label{v3v4fig}}
\end{figure}

Fig.\ \ref{v3v4fig} displays the mean and the width 
of $v_3(N_{part})$ and $v_4(N_{part})$ for Glauber
and KLN initial conditions. While $v_3$ is zero unless event-by-event 
fluctuations are taken into account, $v_4$ is not too different 
between average and event-by-event initial conditions. Please note
that the absolute amount of our $v_3$ seems to be a little larger
than reported but not shown in Ref.\ \cite{Fries:2010jd}.
In both cases, however, even a hypothetical ideal 
experiment capable of determining the impact parameter precisely would be 
unable to distinguish between Glauber and CGC/KLN initial conditions using 
only jet harmonics, since event-by-event physical fluctuations are enough 
to drawn out model sensitivity.

While higher-order coefficients are more sensitive to local gradients, 
they are also more susceptible to event-by-event fluctuations in initial 
conditions (hotspots, etc.), resulting in a larger $v_{3,4}$ event-by-event 
fluctuation.

\begin{figure}[t]
\hspace*{-0.55cm}
\includegraphics[scale=0.5]{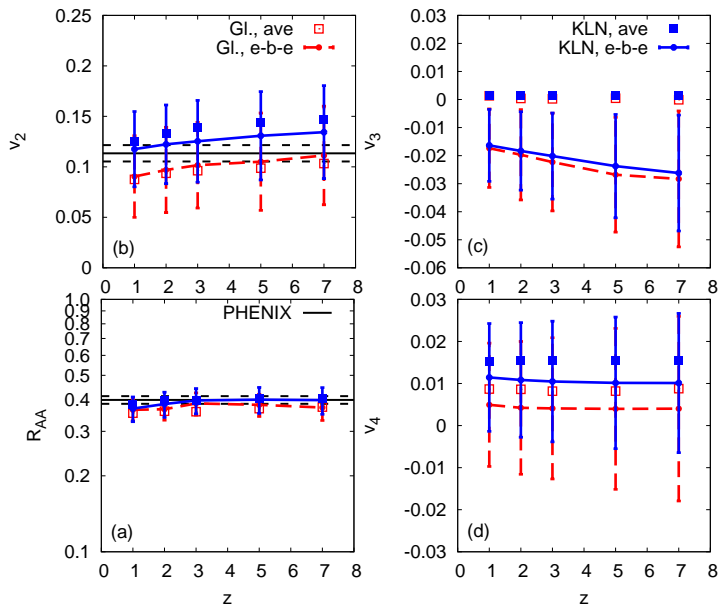}
\caption{(Color online) $R_{AA}$ (a) and $v_n$ (b-d) of high-momentum particles 
at an impact parameter of $b=8$~fm as a function of the
path-length exponent $z$, defined in Eq.\ (\ref{GenericEloss}) for $\tau_0=1$~fm.
PQCD energy loss assumes that $z=1$ \cite{Jia,horojia,renk}, for AdS/CFT on-shell partons $z=2$, 
and AdS/CFT off-shell partons are $z>2$ \cite{ches1,ches2,ches3,Arnold:2011qi}.
A  saturation effect can be seen for all $v_n$ at large $z$. The RHIC data 
(black solid and dashed lines representing the datapoint at $N_{part}=125.7$ and its
errorbar) are taken from Ref.\ \cite{Adare:2010sp}.
\label{expfig}}
\end{figure}

Fig.\ \ref{expfig} shows the sensitivity of $R_{AA}$ and $v_n$ to the 
microscopic mechanism of energy loss, in particular to the power of the
path-length dependence $z$. Here, we consider an impact parameter of $b=8$~fm 
that maximizes $v_n$. As can be seen, once $R_{AA}$ is fixed via the 
coefficient $\kappa$, a residual sensitivity remains mostly in the 
Fourier components $v_2$ and $v_3$. A saturation effect seems to occur
for larger values of $z$.  

Comparing the values of $v_2$ obtained by the Glauber and the KLN model 
in Fig.\ \ref{expfig} to the RHIC data obtained by the PHENIX experiment 
(black solid and dashed lines representing the datapoint at $N_{part}=125.7$ 
and its errorbar) \cite{Adare:2010sp} show that KLN initial conditions
seem to favor lower and Glauber intial conditions higher exponents 
of the path-length dependence. Note that $z>2$ is allowed by AdS/CFT
\cite{ches1,ches2,ches3,Arnold:2011qi}, combined with 
a significant parton virtuality (which is reasonable since the typical 
stopping length is {\em not} $l_f \gg 1/T$). 
Since the virtuality can be directly measured in jet-photon collisions, 
the dependence of the exponent on virtuality could become a quantitative 
signature of AdS/CFT dynamics. Thus, a simultaneus measurement of 
$v_2$, $v_3$, and $v_4$ could elucidate the microscopic mechanism of jet-energy 
loss. 

However, a question that naturally rises in this context is: Why are $v_n$'s at high-$p_T$ so 
insensitive to initial conditions when hydrodynamics shows that the $v_n$'s 
of soft particles are extremely sensitive to initial conditions, leading to 
a systematic error of $\order{100\%}$ in the viscosity \cite{heinzrecent}?
In fact, the difference between these two regimes is {\em not} so 
surprising and can be readily understood physically.
While viscous forces are driven by {\em local gradients} in flow, 
jet absorption is driven by {\em global differences} in the integrated 
$\ave{-\kappa P^a\tau^{z} T^{z-a+2}}$.  The two effects are generally 
not the same and can indeed be very different if the distributions 
(like initial distributions of energy density in a Lorentz-contracted nucleus)
are not smooth.

Both Glauber and KLN initial conditions are tuned to reproduce the observed 
multiplicity distributions, and hence their $\ave{T}$ is similar, 
even if the {\em local gradients} of $T$ might be very different. 
Therefore, if follows that hydrodynamics and tomography lead to
very different results.

In conclusion, we investigated different Fourier harmonics of jet quenching
at RHIC energies and showed that the second Fourier coefficients [$v_2(N_{part})$] are
remarkably sensitive to the initial time $\tau_0$. If this $\tau_0=1$~fm,
as suggested by recent hydrodynamic calculations \cite{heinzrecent},
then the conclusion drawn in Ref.\ \cite{Adare:2010sp} that only CGC/KLN 
initial conditions and an AdS/CFT-like energy loss can simultaneously 
describe the $R_{AA}(N_{part})$ and $v_2(N_{part})$ measured at RHIC 
can no longer be sustained. 
In contrast, for $\tau_0=1$~fm both pQCD-like and AdS/CFT-like energy loss seem to
reproduce the RHIC data well if CGC/KLN initial conditions are taken into account.

Moreover, we studied the microscopic mechanism of jet-energy loss by including an 
energy dependence and exploring the exponent $z$ of the jet path-length 
dependence. We found that higher Fourier harmonics of jet quenching
are remarkably insensitive to the differences between 
the Glauber and CGC/KLN model initial conditions. The 
different $v_n^{I_{AA}}(N_{part})$ vs.\ $v_n(N_{part})$ 
correlations between the moments of monojet 
and dijet nuclear modifications factors remain a very sensitive 
probe to differentiate between Glauber and CGC/KLN initial conditions.

\section*{Acknowledgments}
B.B.\ is supported by the Alexander von Humboldt foundation via a Feodor 
Lynen fellowship. M.G. and B.B.\ acknowledge support from DOE under
Grant No.\ DE-FG02-93ER40764.
G.T. acknowledges the financial support received from the Helmholtz 
International Center for FAIR within the framework of the LOEWE program
(Landesoffensive zur Entwicklung Wissenschaftlich-\"Okonomischer
Exzellenz) launched by the State of Hesse. The authors thank A.\ Dumitru
for providing his KLN code to simulate the CGC initial conditions.

\end{document}